\documentclass[prb,twocolumn,superscriptaddress,showpacs,preprintnumbers,amsmath,amssymb]{revtex4}

\usepackage{graphicx}
\usepackage{dcolumn}
\usepackage{bm}
\usepackage{SIunits}
\usepackage{tabularx}
\usepackage{threeparttable}

\begin{document}
\title{Structural and vibrational properties of phenanthrene under pressure }
\author{Qiao-Wei Huang}
\affiliation{Department of Physics, South China University of Technology, Guangzhou 510640, China}

\author{Jiang Zhang}
\affiliation{Department of Physics, South China University of Technology, Guangzhou 510640, China}

\author{Adam Berlie}
\affiliation{Center for Energy Matter in Extreme Environments and Key Laboratory of Materials Physics, Institute of Solid State Physics, Chinese Academy of Sciences, Hefei 230031, China}

\author{Zhen-Xing Qin}
\affiliation{Department of Physics, South China University of Technology, Guangzhou 510640, China}

\author{Xiao-Miao Zhao}
\affiliation{Department of Physics, South China University of Technology, Guangzhou 510640, China}

\author{Jian-Bo Zhang}
\affiliation{Department of Physics, South China University of Technology, Guangzhou 510640, China}

\author{Ling-Yun Tang}
\affiliation{Department of Physics, South China University of Technology, Guangzhou 510640, China}

\author{Jing Liu}
\affiliation{Institute of High Energy Physics, Chinese Academy of Sciences, Beijing 100190, China}

\author{Chao Zhang}
\affiliation{Department of Physics, Yantai University, Yantai 264005, China}
\affiliation{Beijing Computational Science Research Center, Beijing 100084, China}

\author{Guo-Hua Zhong}
\affiliation{Shenzhen Institutes of Advanced Technology, Chinese Academy of Sciences, Shenzhen 518055, China}

\author{Hai-Qing Lin}
\affiliation{Beijing Computational Science Research Center, Beijing 100084, China}
\affiliation{Shenzhen Institutes of Advanced Technology, Chinese Academy of Sciences, Shenzhen 518055, China}

\author{Xiao-Jia Chen}
\email{xjchen@ciw.edu}
\affiliation{Geophysical Laboratory, Carnegie Institution of Washington, Washington, DC 20015, USA}
\affiliation{Department of Physics, South China University of Technology, Guangzhou 510640, China}

\begin{abstract}

The structural and vibrational properties of phenanthrene are measured at high pressures up to 30.2 GPa by Raman spectroscopy and synchrotron x-ray diffraction techniques. Two phase transitions are observed in raman spectra under pressures of 2.3 GPa and 5.4 GPa with significant changes of intermolecular and intramolecular vibrational modes, and all the raman peaks disappeared in the luminesce background above 10.2 GPa. Upon further compression above 20.0 GPa, three broad bands are observed respectively around 1600, 2993 and 3181 cm$^{-1}$ in Raman spectra, strongly indicating that phenanthrene has transformed into amorphous hydrogenated carbon. Based on x-ray diffraction, the structures of corresponding phases observed in Raman spectra are  identified with space group of $P_{\rm}$2$_{1}$ for phase I at 0-2.2 GPa, $P_{\rm}$2/\emph{m} for phase II at 2.2-5.6 GPa and $P_{\rm}$2/\emph{m}+\emph{Pmmm} for phase  III at 5.6-11.4 GPa , and the structure above 11.4 GPa is identified with space group of \emph{Pmmm}. Although phenanthrene has amorphized above 20.0 GPa, the amorphous hydrogenated carbon clusters still remain the crystalline characteristic based on x-ray diffraction patterns. Our results suggest that the long-range periodicity and the local disorder state coexist in phenanthrene at high pressures.

\end{abstract}


\vskip 300 pt

\maketitle

\section{INTRODUCTION}

Great interests are attracted on the studies of the distinct properties of aromatic hydrocarbon, such as specific electronic, optoelectronic, and optical applications.\cite{pah1,pah2,pah3} Recently, a series of aromatic hydrocarbons such as phenanthrene\cite{XFWang,XFWang1} (C$_{14}$H$_{10}$), picene\cite{Mitsuhashi} (C$_{22}$H$_{14}$), coronene\cite{Kubozono} (C$_{24}$H$_{12}$), and 1,2:8,9-dibenzopentacene\cite{GFChen} (C$_{30}$H$_{18}$) are found to exhibit superconductivity at temperatures from 5 to 33 K by doping alkali or alkali-earth metal. In contrast to the previous organic superconductors, such as tetrathiafulvalene\cite{TTF} (TTF), bis-ethylenedithrio-TTF\cite{BEDTTF} (BEDT-TTF, abbreviated as ET), and tetramethyltetraselenafulvalene\cite{TMTSF} (TMTSF) derivatives, hydrocarbon superconductors are a novel superconducting series with intriguing properties. Based on previous discoveries, $T_{\rm c}$ of hydrocarbon superconductor is enhanced with the increasing number of benzene ring\cite{XFWang,Mitsuhashi,Kubozono,GFChen}, reaching the highest $T_{\rm c}$=33.1 K in organic superconductor comparable to the carbon based superconductor metal-doped fullerides\cite{C60} ($T_{\rm c}$=38 K) under pressure. The pressure effect of such system is remarkable that $T_{\rm c}$ is nearly doubled in the 18 K phase of K$_{3}$picene from 18 to 30 K under 1.2 GPa.\cite{Kubozono} This positive pressure effect is also observed in A$_x$phenanthrene (A=K,Rb, Ba, Sr) under pressure.\cite{XFWang,XFWang1} For the 7 K phase of K$_{3}$picene, however, $T_c$ decreases slightly upon compression up to 1.0 GPa\cite{Kubozono}, suggesting the complex mechanism in such superconducting system. Besides, the feature of ``armchair" edge type, which is regarded as a key factor of superconductivity in such a system, is shared in the hydrocarbon superconductors in contrast to the ``zigzag" edge type.\cite{Kato,XFWang,Mitsuhashi,GFChen} Hydrocarbon superconductors are also a low-dimensional system with strong electron-phonon and electron-electron interactions, and both of which may be involved in superconductivity as suggested by some theoretical works.\cite{Kato2,Subedi,Casula,Giovannetti} But the mechanism of superconductivity is still beyond understanding in hydrocarbon superconductor due to the complexity in such a system. Thus, more detailed experiments are required to investigate the characteristic of hydrocarbon superconductor, also in undoped state, in order to thoroughly understand the mechanism of superconductivity.

Compression is a useful way to investigate the properties of superconductor, also in their undoped state. For undoped sample, application of external pressure can alter the distribution of charges, possibly leading to the metallic state which may have superconducting properties.\cite{Witting,Bundy,Struzhkin,Shimizu} Previous investigations revealed that pentancene, which is the isomer of picene with ``zigzag" edge type, can be transformed into metal state under 27 GPa.\cite{Aust} In addition, applied pressure can modify the structure of sample, and possibly induce the coupling of vibration modes and the wave functions overlap\cite{Strobel,XJC,Kung}, which is important to the properties of superconducting state. The high pressure behaviors of most aromatic hydrocarbons are still absent and call for further investigations.

In this paper, we present a high-pressure investigation of phenanthrene by combining raman scattering and synchrotron x-ray diffraction (XRD) techniques up to 30.8 GPa. We choose phenanthrene for our studies because it is the smallest and simplest molecule in the hydrocarbon family with ``armchair" edge type molecular structure, and much larger superconducting volume faction was reported in cation doped phenanthrene than other hydrocarbon superconductors.\cite{XFWang1} In our studies, three possible phases exist in phenanthrene up to 30.8 GPa, which are identified by Raman scattering and synchrotron XRD measurements. The corresponding structure of each phase is determined based on the obtained synchrotron XRD data. Besides, our results suggest that phenanthrene becomes amorphous hydrogenated carbon above 20.0 GPa based on Raman spectra, but XRD patterns suggest that the structure of the compound still remains above 20.0 GPa.

\section{EXPERIMENTAL DETAILS}

Phenanthrene of 98\% purity, colorless crystal, was purchased from TCI Co. and used without any further purification. High-pressure Raman measurements were carried out in a symmetric diamond anvil cells (DACs) with the culets of 300 microns. A 100 microns hole which was drilled in the center of stainless-steel gasket served as the sample chamber. One small piece of the sample was placed in the chamber and small ruby grains were put on one of the diamond culets that served as the pressure calibration. The pressure acting on the sample can be determined by the shift of the ruby wave number by the well-established ruby fluorescence method.\cite{HKMao} No pressure transmitting medium was used for the measurements. Renishaw Invia Raman system with a spectrometer (1800 lines/mm grating) was used for the measurement, giving a resolution of 1 cm$^{-1}$. The Raman spectra was measured in backscattering geometry with visible laser excitation (532 nm) with power less than 50 mW. The spectra were collected from 100 to 3300 cm$^{-1}$.

\begin{figure}[tbp]
\includegraphics[width=\columnwidth]{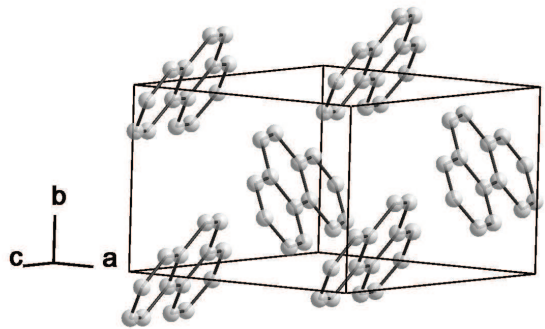}
\caption{Molecular structure and crystal structure of phenanthrene at ambient pressure.}
\end{figure}

High-pressure synchrotron XRD experiments were performed with the same DAC at room temperature, using synchrotron radiation at the Beijing Synchrotron Radiation Facility (BSRF). The wavelength of the x-ray is 0.6199 {\AA}. The sample-to-detector distance and the image plate orientation angles were calibrated using CeO$_{2}$ standard. No pressure transmitting medium was used for the measurements. The two-dimensional diffraction images were converted to 2$\theta$ versus intensity data plots using the FIT2D software.

\section{RESULTS AND DISCUSSION}

\subsection{Evolution of vibrational properties with pressure}

The crystal structure and molecular structure of phenanthrene is shown in Fig. 1. The crystal belongs to the space group $P_{\rm}$2$_{1}$ with two molecules in the unit cell. The lattice parameters are $a$=8.472(4), $b$=6.166(4), $c$=9.467(5){\AA}, and $\beta$=98.01$\degree$ at ambient pressure.\cite{Trotter} Because the molecule of phenanthrene consists of 24 atoms, there are 66 internal vibration modes predicted on the basis of the molecular vibration theory. There are 45 in-plane and 21 out-of-plane normal modes of vibration in phenanthrene spanning irreducible representations as 23A$_{1}$+22B$_{2}$+11A$_{2}$+10B$_{1}$.\cite{Bandy} According to the selection rule, A$_{1}$, B$_{2}$, A$_{2}$ and B$_{1}$ are raman active. But some of these modes can not be detected in our experiment due to the weak intensity. The vibrations can be divided into two different types in raman spectra: intermolecular modes and intramolecular modes.\cite{Teixeira} The vibrations of intermolecular modes, representing essentially the rotations and translations of rigid molecules, are in the low frequency region of the Raman spectra, while the other vibrations in the high frequency region are represented the intramolecular modes. Table I lists the most intense peaks in spectra and the assignments of their vibration types reported previously by Bree et al.\cite{Bree}. The vibrations of phenanthrene can be mainly divided into five types: lattice vibration, CH out of plane vibration, CC stretching+CH bending vibration, CC stretching vibration, and CH stretching vibration.

\begin{figure}[tbp]
\includegraphics[width=\columnwidth]{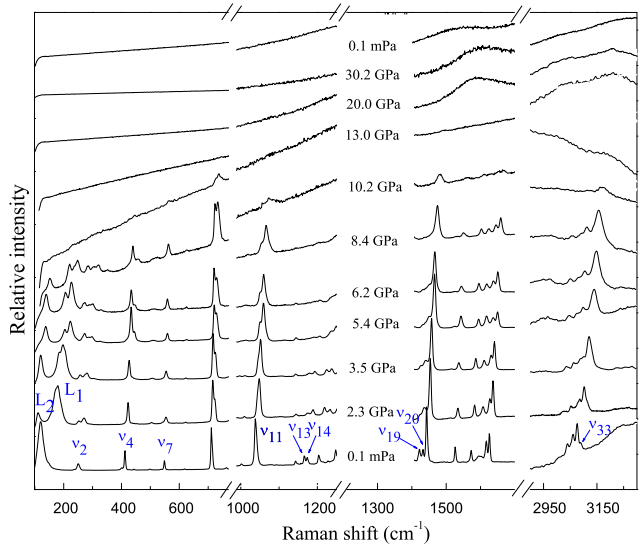}
\caption{(color online) Representative Raman spectra of phenanthrene in the full spectral regions at ambient conditions upon compression to 30.2 GPa.}
\end{figure}

Raman spectras of phenanthrene are collected from ambient pressure to high pressure up to 30.2 GPa followed by decompression, and the selected spectra in the frequency region of 100-3300 cm$^{-1}$ are shown in Fig. 2. With increasing pressure, all the Raman peaks shift to higher frequencies except a peak originally at 1526 cm$^{-1}$. Unpond further compression, several peaks become weak and vanished below 10.2 GPa, and the other Raman peaks are hard to be observed relative to the inreasing luminescence background above 10.2 GPa. When pressure is further increased above 20.0 GPa, the luminescence background vanishes but no original Raman peaks reappear. Meanwhile, three broad bands appear respectively around 1600, 2993 and 3181 cm$^{-1}$. The broad bands shift sluggishly to higher frequencies upon further compression, and the broad band still remained upon the release of pressure while the original Raman modes are not recovered.

\begin{table}[ht]
\centering
\caption{Intramolecular vibrations of phenanthrene:Vibrational aasignment, symmetry and observed frequency.}
\begin{ruledtabular}
\renewcommand\arraystretch{1.25}
\begin{tabular}{c c c c}
Modes & Assignment$^{a,b}$ & Symmetry & Obs (cm$^{-1}$)\\ \hline
$\nu$$_{1}$ & $\alpha$(CCC) & B$_{2}$ & 236\\
$\nu$$_{2}$ & $\delta$(CCCC) & A$_{1}$ & 251\\
$\nu$$_{3}$ & $\delta$(CCCC) & A$_{1}$ & 401\\
$\nu$$_{4}$ & $\alpha$(CCC) & B$_{2}$ & 412\\
$\nu$$_{5}$ & $\delta$(CCCC) & B$_{1}$ & 443\\
$\nu$$_{6}$ & $\alpha$(CCC) & B$_{2}$ & 498\\
$\nu$$_{7}$ &             & B$_{1}$ & 537\\
$\nu$$_{8}$ & $\alpha$(CCC) & A$_{1}$ & 549\\
$\nu$$_{9}$ & $\gamma$(HCCC)& A$_{1}$ & 710\\
$\nu$$_{10}$ & $\alpha$(CCC) & B$_{1}$ & 713\\
$\nu$$_{11}$ & t(CC) & A$_{1}$ & 1039\\
$\nu$$_{12}$ & t(CC)+$\beta$(HCC) & A$_{1}$ & 1144\\
$\nu$$_{13}$ & t(CC)+$\beta$(HCC) & A$_{1}$ & 1166\\
$\nu$$_{14}$ & $\delta$(CCCC) & A$_{1}$ & 1173\\
$\nu$$_{15}$ & $\delta$(CCCC) & A$_{1}$ & 1204\\
$\nu$$_{16}$ & 1348a-115b=1233b & B$_{1}$ & 1228\\
$\nu$$_{17}$ & $\beta$(HCC) & B$_{1}$ & 1248\\
$\nu$$_{18}$ & t(CC)+$\beta$(HCC) & A$_{1}$ & 1353\\
$\nu$$_{19}$ & 710a$\times$2=1420a & A$_{1}$ & 1422\\
$\nu$$_{20}$ & t(CC)+$\beta$(HCC) & A$_{1}$ & 1432\\
$\nu$$_{21}$ & t(CC)+$\beta$(HCC) & A$_{1}$ & 1443\\
$\nu$$_{22}$ & t(CC) & A$_{1}$ & 1526\\
$\nu$$_{23}$ & t(CC) & B$_{1}$ & 1573\\
$\nu$$_{24}$ & t(CC) & A$_{1}$ & 1596\\
$\nu$$_{25}$ & t(CC) & B$_{1}$ & 1617\\
$\nu$$_{26}$ & t(CC) & A$_{1}$ & 1626\\

\end {tabular}
\end{ruledtabular}
\footnotetext[1]{Internal coordinate are defined as follows: t is C-C stretching, $\alpha$ is CCC bending, $\beta$ is HCC bending, $\gamma$ is HCCC out-of-plane bending, $\delta$ is CCCC torsion.}
\footnotetext[2]{From Ref. [30].}
\end{table}

According to the previous vibrational assignments\cite{Bree}, the intermolecular modes of phenanthrene occupies up to 200 cm$^{-1}$, but only one intermolecular mode can be observed at ambient pressure in our measurement. The intermolecular mode at 143 cm$^{-1}$ appears at 2.3 GPa, which is originally at the frequency range below 100 cm$^{-1}$ at ambient pressure. Additionally, a new mode appears at the shoulder of intermolecular mode L${_1}$ at 2.3 GPa, and splitted into two modes with increasing pressure. The splitting in intermolecular mode can be identified as the response of structural transformation with application of external pressure. Furthermore, one C-H stretching mode $\nu$$_{33}$ originally at 3088 cm$^{-1}$ disappeared at 2.3 GPa. With increasing pressure to 5.4 GPa, the Raman spectra changes dramatically, indicating substantial changes in the crystal and/or molecular structures. As shown in Fig. 2, $\nu$$_{2}$ with the CCCC torsion, $\nu$$_{4}$ with CCCC bending and $\nu$$_{11}$ with CC stretching respectively splited into two modes above 5.4 GPa. Additionally, several intramolecular modes  $\nu$$_{13}$, $\nu$$_{14}$, $\nu$$_{19}$ and $\nu$$_{20}$ vanished at pressures above 5.4 GPa. Compressing continually, all the peaks gradually become board and disappear in the increasing luminescence background around 10.2 GPa. In frequencies range from 1500 to 1700 cm$^{-1}$, no apparent discontinuity, disappearance, or splitting was observed in the C-C stretching vibration modes under pressure, implying the stability of carbon bonding upon compression.

\begin{figure}[tbp]
\includegraphics[width=\columnwidth]{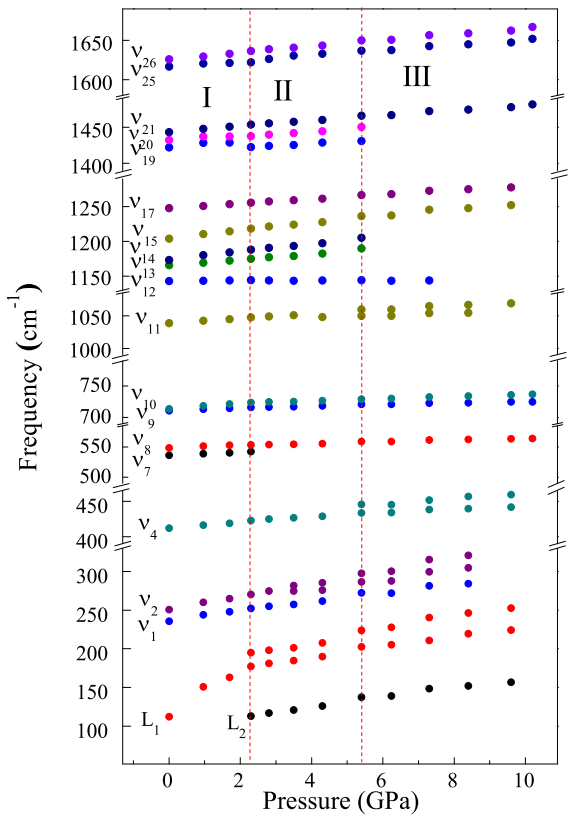}
\caption{(color online) Pressure dependence of the frequencies of phenanthrene for the observed modes in all regions at room temperature. The vertical dashed lines at near 2.3 and 5.4 GPa indicate the proposed phase boundaries.}
\end{figure}

The selected pressure dependence of the Raman modes in 100-1700 cm$^{-1}$  was depicted in Fig. 3 to reveal the possible phase transitions under pressure. There are three phases existed in phenanthrene under pressure labeled by phase I (0-2.3 GPa), phase II (2.3-5.6 GPa), and phase III (5.6-10.2 GPa). Upon compression from ambient pressure to 2.3 GPa, the crystalline structural transition can be identified by several features of Raman spectra. As seen in Fig. 3, the slope of mode L$_{1}$ drastically decreased from 27.7 to 6.6 cm$^{-1}$/GPa around 2.3 GPa indicating that the crystal structure of phenanthrene transformed to the phase II. The molecules of phenanthrene were forced to be rearranged to a more compact configuration to adapt the external stress. Besides, the intermolecular mode L$_{1}$ at 196 cm$^{-1}$ splits into two modes at 2.3 GPa giving further evidence of structural phase transition.  In intramolecular range, it is also noted that the mode $\nu$$_{7}$ and C-H stretching mode $\nu$$_{33}$ vanished at pressure of 2.3 GPa. Thus, we suggested that the phase transition at 2.3 GPa should be associated with the modification of the molecular configuration. The structural phase transition is consistent with the previous study of fluorescence spectra on phenanthrene upon compression\cite{Jones}, in which the pronounced decrease in total intensity was observed above 2.5 GPa. Compressing continually, obvious changes in intramolecular modes at 5.4 GPa indicates that the phase II starts to transformed to phase III. All the Raman modes disappeared above 10.2 GPa due to the luminescence background, possibly indicating another phase transition.

Although no Raman peaks is observed above 10.2 GPa, the Raman spectra measurement was continued to be perform upon compression up to 30.2 GPa because no direct evidence was found to illustrate that phenanthrene undergoes metallization (i.e., visible darkening of the sample in DAC) under pressure up to 10.2 GPa. Upon further compression, luminescence background disappeared, but no original Raman peaks recover. However, a broad band is observed in the range of 1500-1700 cm$^{-1}$ above 20.0 GPa. The band corresponds to the stretching motion of $sp^{2}$ carbon pairs in the rings or chains, which is an important characteristic of disordered, nanocrystalline and amorphous carbon\cite{Jackson,Ferrari}. This suggests that  phenanthrene has amorphized at high pressures. Additionally, we also observed another two broad bands around 2993 and 3181 cm$^{-1}$ above 20.0 GPa. These two bands could be assigned respectively to the C-H stretching vibration and the second order mode of $sp^{2}$ carbon pairs stretching.\cite{Jackson,Ferrari} The C-H characteristic indicated the fact that the phenanthrene compound transformed to amorphous hydrogenated carbon instead of carbonization reported by the high pressure shock waves studies on phenanthrene\cite{Mimurra}. Upon further compression, the position of broad bands just changes slightly toward higher frequency, indicating that amorphous network are superhard. The broad bands originating from amorphous network moved back to lower frequencies upon followed decompression.  The result is similar to the observations of other aromatic compound that compressed pyrene and compressed benzene also transformed to aromatic hydrogencarbon at high pressures.\cite{sun,Ciabini1,Ciabini2,Ciabini3}

\begin{figure}[tbp]
\includegraphics[width=\columnwidth]{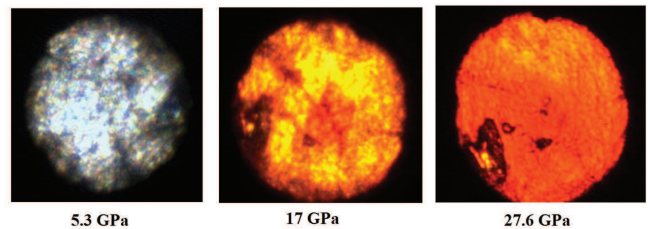}
\caption{(color online) Photos of phenanthrene under back illumination at pressure of 5.3, 17 and 27.6 GPa. The sample gradually changed from colorless to darken but still keep visible upon compression. }
\end{figure}

In addition, we recorded the changes of colors of phenanthrene with increasing pressure (as shown in Fig. 4). The sample gradually changed from colorless, yellow and finally dark red. This color change can be explained by a red shift of the absorption peaks as a function of pressure which was already experimentally observed\cite{Wied} as well as theoretically considered.\cite{wein} The sample still keep visible under back illumination upon compression up to 30.0 GPa. It indicated that phenanthrene does not metalize or carbonize under pressure.

\begin{figure}[tbp]
\includegraphics[width=\columnwidth]{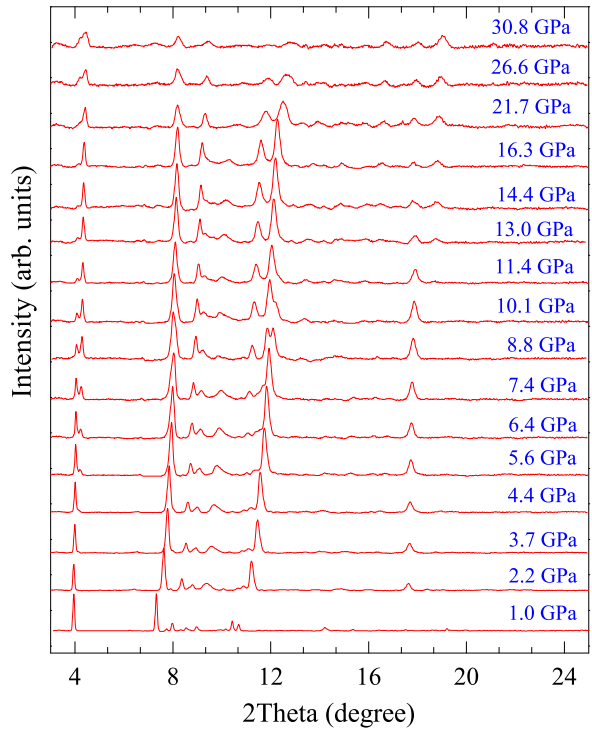}
\caption{(color online) Synchrotron x-ray ($\lambda$= 0.6199 {\AA}) diffraction patterns of phenanthrene during the pressurization from ambient conditions to 30.0 GPa. The background curves have been removed manually.}\label{fig4}
\end{figure}

\subsection{Analysis of structural evolution with pressure}

The XRD of phenanthrene were collected from ambient pressure to 30.8 GPa (Fig. 5). The diffraction peaks shift to larger angles indicating the shrinkage of the phenanthrene lattice. XRD pattern at 2.2 GPa shows that phenanthrene has transformed to another phase on the basic of distinction from ambient pressure pattern, as found by our Raman measurements. Upon further compression, a new reflection peak appears around the (110) peak at 4.0$^{o}$ at 5.6 GPa, and gradually strengthen accompanied with the decreasing intensity of the (110) peak upon compression. The corresponding changes were also observed in the Raman spectra at 5.4 GPa, strongly indicating a structural phase transitions in phenanthrene. Compressing continually to 8.8 GPa, the peak at 12.0$^{o}$ splits and becomes two peaks with increasing pressure, and the latter reflection peak becomes weaker and disappears around 11.4 GPa. At higher pressure, another new peak appears at 18.6$^{o}$ above 11.4 GPa. All the reflections become weaker, but still remain up to 30.8 GPa.

\begin{figure}[tbp]
\includegraphics[width=\columnwidth]{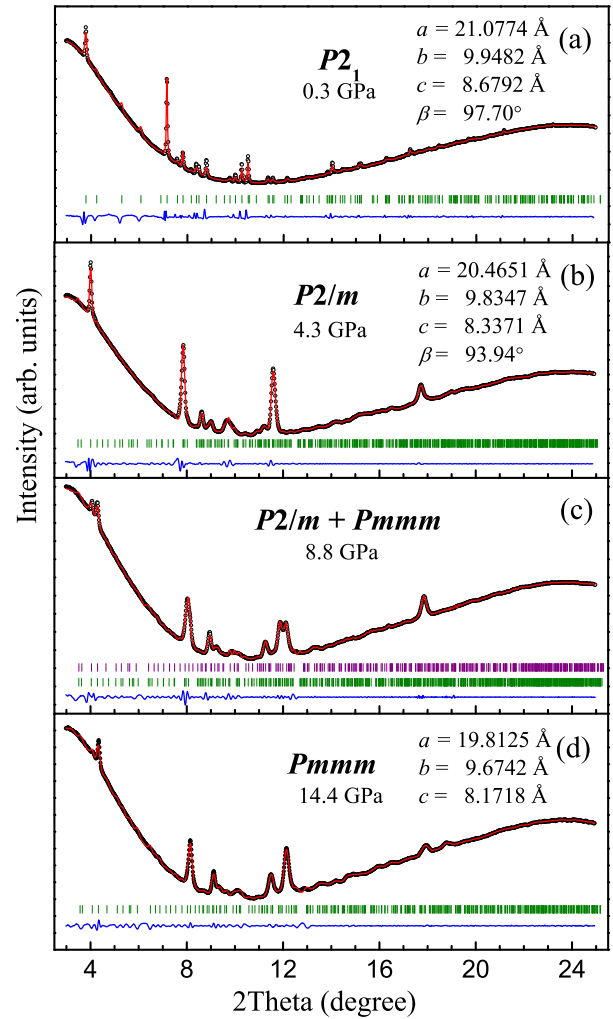}
\caption{(color online) X-ray diffraction patterns of phenanthrene at pressures of 0.3 (a), 4.3 (b), 8.8 (c) and 14.4 (d) GPa. The refined lattice parameters for the corresponding space groups are given respectively. The open circles represent the measured intensities and the red lines the results of profile refinements by the best LeBail-fit with each space group. The positions of the Bragg reflections are marked by vertical lines and the difference profiles are shown at the bottoms (blue lines). }
\end{figure}

The profile fitting of the patterns was performed by the Le-Bail method using the GSAS program\cite{GSAS} to investigate the crystal structure at different pressures. The fitted results at selected pressure are shown in Fig. 6. The first phase is refined on the basis of parameters of the known single-crystal structure crystallizing in the space group \emph{P}2$_{1}$. The volume decreased 8.6\% across the first phase according to the result of fitting. The first phase transition is determined around 2.2 GPa according to obvious changes in diffraction patterns. However, the details of crystal structures were still unknown at the second phase. Thus, we try to use the program Dicovol06 and Peakfit v4 to investigate the possible crystal structure available at higher pressures. For phase II, all the diffraction peaks were resolved at 2.2 GPa and indexed mainly to the monoclinic system or orthorhombic system. Compared to the similar organic compound coronene\cite{coronene}, investigations reveled that coronene transformed to the body centered orthorhombic structure from monoclinic phase below 2.0 GPa. However, the orthorhombic phase group is not suitable in our cases because of the bad fitting results. Therefore, the monoclinic space group of $P2/m$ is selected as the candidate for refinement of phase II due to reasonable figures of merit (M,F) and/or volume of the cell. The lattice parameters of phase II are $a$=21.139(9), $b$=9.947(8), $c$=8.531(7) {\AA} and $\beta$=93.866(3)$\degree$ at 2.2 GPa. Based on the refinements, the predicted volume per formula unit of phenanthrene is 223.75 {\AA}$^{3}$ at 2.2 GPa assuming Z=8. From 5.6 to 11.4 GPa, all the diffraction peaks could still be fitted based on the unit cell of space group $P_{\rm}$2/$m_{\rm}$ except the additional peaks appearing above 5.6 GPa. This indicates that the structure of phase II either becomes more complex or coexists with an appearing new high-pressure phase. Upon further compression above 11.4 GPa, the two additional peaks have completely been replaced the original reflection peaks of the second phase at 4.1$^{o}$ and 12.1$^{o}$, suggesting the intermediate phase is completed above 11.4 GPa. This intermediate phase is consistent with the phase III observed in the Raman spectra. For determining the additional phase in the intermediate phase, we should first figure out the structural phase above 11.4 GPa. According to the Raman spectra, the number of vibrational modes decreases above 5.6 GPa, indicating the new phase with higher symmetry. Compared to the results of coronene, phenanthrene likely changed into the orthorhombic structure. Therefore, we choose orthorhombic space group \emph{Pmmm} to fit the profile of XRD above 11.4 GPa based on the indexed results. For intermediate phase, additionally, only \emph{Pmmm} and \emph{P}2/\emph{m} are selected to fit the pattern at 8.8 GPa, as shown in Fig. 6(c).

Using the GSAS software, we obtain the lattice parameters from the structural fitting. The values of volume per formula unit can be well derived from the lattice parameters. Their variations are important references of the compressibility of each phase. The pressure dependence of the volume per formula unit is showed in Fig. 7, and the dependence can be decided systematically by an appropriate equation of state (EoS). For phenanthrene, the Birch-Murnaghan equation of state (BM3 EoS)\cite{Brich} is used to obtain the bulk modulus and its derivative. The Birch-Murnaghan equation of state is defined as
\begin{equation}
P=3K_0f_E(1+2f_E)^{\frac{5}{2}}[1+\frac{3}{2}(K_0^{'}-4)f_E]
\end{equation}
where $f_E=[(\frac{v_0}{v})^{\frac{2}{3}}-1]$, $V_0$ is the volume per formula unit at ambient pressure, $V$ is the volume at pressure $P$ given in GPa, $K_0$ is the bulk modulus at ambient pressure, and ${K_0'}$ is value of the first pressure derivative of bulk modulus at ambient pressure. $K_0$ and $K_0^{'}$ are independent to each other. The first phase are not be fitted with EoS as limited data, and the lines are the fit to the data points of other two phases above 2.2 GPa. The bulk moduli of phase \emph{P}2/\emph{m} is 16.16 GPa with $K{_0'}$=28.43, $V{_0}$=235.57 {\AA}$^{3}$ and phase \emph{Pmmm} is 19.45 GPa with  $K{_0'}$=19.91, $V{_0}$=237.1684 {\AA}$^{3}$,respectively. The bulk moduli of phenanthrene above 2.2 GPa is nearly twice more than the ambient bulk moduli of anthrance\cite{Oehzelt1,Oehzelt2} which is the isomer of phenanthrene. The increasing bulk moduli infers an enhancement of bond strength during phase transitions and indicates the intrinsic higher compressibility.

\begin{figure}[tbp]
\includegraphics[width=\columnwidth]{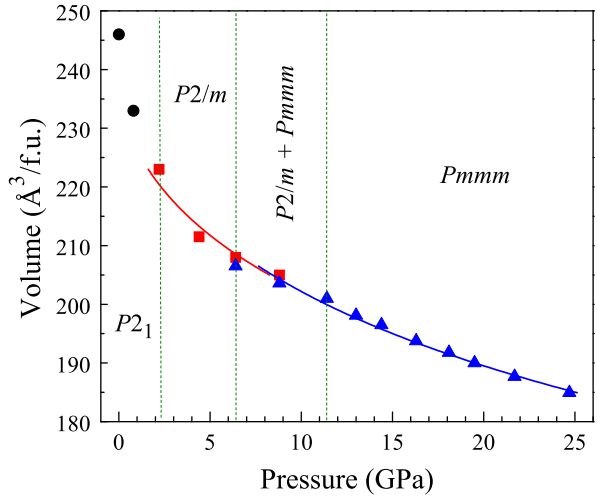}
\caption{(color online) Volume per formula unit change of phenanthrene with pressure. The solid lines demonstrate the fitting data of phases to the Birch-Murnaghan equation of state and vertical dashed lines denote the phase boundaries.}
\end{figure}

Additionally, an intriguing feature is observed in phenanthrene at high pressures based on the combing Raman spectra and XRD patterns. As shown in Raman spectroscopy, the broad bands observed above 20.0 GPa strongly indicate that phenanthrene has transformed to amorphous hydrogenated carbon at high pressures, but the long-range structural order is still exited in phenanthrene up to 30.8 GPa based on the XRD patterns. The observations strongly suggest that inherent disorder are present in a crystal. Although the molecular structure of phenanthrene has been broken down at high pressures, the amorphous cluster likely stay on the lattice sites and remain the periodic structure under high pressure. The similar observation is also observed in solvated C$_{60}$ at high pressure reported by Wang $et$ $al$.\cite{LWang} The C$_{60}$ molecules from the crystalline solvated fullerene phase C$_{60}$*m-xylene undergoes an order-to-disorder transition under compression around 35 GPa but keeps their translational symmetry. The result is in contrast to the pure C$_{60}$, where the face-centered cubic periodicity of the C$_{60}$ molecular units disappears when it amorphizes above 30 GPa. Thus, they suggested that the m-xylene solvent molecules play a crucial role in maintaining the long-range periodicity in the ordered amorphous carbon clusters. In our experiments, coexistence of order and disorder state is obtained in the pure phenanthrene without spacer molecules in contrast to C$_{60}$*m-xylene. The certain strong interaction must be existed in such system maintaining the long-range order at high pressures. This is important to develop the potential physical properties in this system, such like that the potentially higher $T_C$ of superconductivity could be existed in doped phenanthrene upon heavy compression due to the stability of crystalline characteristic under pressure.
\section{conclusions}

We performed high pressure measurements of Raman scattering and synchrotron x-ray diffraction of phenanthrene up to 30.8 GPa at room temperature. We determined three phases existed in phenanthrene up to 30.8 GPa. The space groups for the phases were identified to be as follows: $P_{\rm}$2$_{1}$ at 0-2.2 GPa, $P_{\rm}$2/\emph{m} at 2.2-5.6 GPa, and \emph{Pmmm} at 11.4-30.8 GPa. Besides, phenanthrene was found to turn into the amorphous hydrogenated carbon clusters with long range periodicity above 20.0 GPa.

\begin{acknowledgments}
This work was supported as part of EFree, an Energy Frontier Research Center funded by the U.S. Department of Energy (DOE), Office of Science under
DE-SC0001057. The work done in China was supported by the Cultivation Fund of the Key Scientific and Technical Innovation Project Ministry of Education of China (No.708070), the Shenzhen Basic Research Grant (No. JC201105190880A), the National Natural Science Foundation of China (No. 11274335), Guangdong Natural Science Foundation (No. S2012040007929), and the Fundamental Research Funds for the Central Universities SCUT (No.2012zz0078).
\end{acknowledgments}

\end{document}